\documentclass[aps,preprintnumbers,superscriptaddress,longbibliography,amsmath,amssymb,twocolumn,10pt,floatfix]{revtex4-2}

\usepackage{graphicx}
\usepackage{dcolumn}
\usepackage{bm}
\usepackage{xspace}
\usepackage{multirow}
\usepackage{booktabs}
\usepackage{xcolor}
\usepackage{xfrac}
\usepackage[caption=false]{subfig}
\graphicspath{{./plots/}}

\usepackage{hyperref}
\hypersetup{
    colorlinks=true,
    linkcolor=blue,
    filecolor=blue,      
    urlcolor=blue,
    pdftitle={Overleaf Example},
    pdfpagemode=FullScreen,
    }

\newcommand{\Nmu}{N_\mu}
\newcommand{\Smu}{S_\mu}

\def\qgs{QGSJet\,II-04\xspace}

\usepackage{units}
\usepackage{comment}

\newcommand{\x}[1]{%
  {}$
  \kern-2\mathsurround 
  $
  \binoppenalty10000 \relpenalty10000 #1
  {}$
  \kern-2\mathsurround 
  $
}

\begin{document}


\title{Azimuthal fluctuations and number of muons at the ground in muon-depleted proton air showers at PeV energies}
\date{\today}

\author{A.~Bakalov\'a}
\affiliation{Institute of Physics of the Czech Academy of Sciences, Prague, Czech Republic}

\author{R.~Concei\c{c}\~ao}
\affiliation{LIP - Laborat\'orio de Instrumenta\c{c}\~ao e F\'isica Experimental de Part\'iculas, Lisbon, Portugal}
\affiliation{Departamento de F\'isica, Instituto Superior T\'{e}cnico, Universidade de Lisboa, Lisbon, Portugal}

\author{L.~Gibilisco}
\affiliation{LIP - Laborat\'orio de Instrumenta\c{c}\~ao e F\'isica Experimental de Part\'iculas, Lisbon, Portugal}
\affiliation{Departamento de F\'isica, Instituto Superior T\'{e}cnico, Universidade de Lisboa, Lisbon, Portugal}

\author{V.~Novotn\'y}
\affiliation{Institute of Physics of the Czech Academy of Sciences, Prague, Czech Republic}
\affiliation{Charles University, Faculty of Mathematics and Physics, Institute of Particle and Nuclear Physics, Prague, Czech Republic}

\author{M.~Pimenta}
\affiliation{LIP - Laborat\'orio de Instrumenta\c{c}\~ao e F\'isica Experimental de Part\'iculas, Lisbon, Portugal}
\affiliation{Departamento de F\'isica, Instituto Superior T\'{e}cnico, Universidade de Lisboa, Lisbon, Portugal}

\author{B.~Tom\'e}
\affiliation{LIP - Laborat\'orio de Instrumenta\c{c}\~ao e F\'isica Experimental de Part\'iculas, Lisbon, Portugal}
\affiliation{Departamento de F\'isica, Instituto Superior T\'{e}cnico, Universidade de Lisboa, Lisbon, Portugal}

\author{J.~V\'icha}
\affiliation{Institute of Physics of the Czech Academy of Sciences, Prague, Czech Republic}

\begin{abstract}
    Muon counting is an effective strategy for discriminating between gamma and hadron-initiated air showers. However, their detection, which requires shielded detectors,  is highly expensive and challenging to implement across large, environmentally sensitive areas.
    This work allowed to establish for the first time that at PeV energies the gamma/hadron discriminator based on the new $LCm$ variable have proton rejection levels of the order of $10^{-4}$, outperforming the discrimination power based on the counting of the number of muons.
    A thorough examination of muon depleted showers at the PeV energies and the simulation strategy devised to achieve the required $\mathcal{O}(10^6)$ simulated showers is presented.

\end{abstract}

\pacs{Valid PACS appear here}
\maketitle


\section{Introduction}

The direct detection of the number of muons at ground level ($N_{\mu}$) is widely regarded as the most effective method to achieve very high rejection factors for gamma/hadron discrimination (around $10^4-10^5$) at PeV energies. This approach was successfully implemented by the LHAASO collaboration~\cite{LHAASO_PeV}, leading to the discovery of the first PeV gamma-ray sources in our Galaxy, opening a new exciting and unexpected chapter in the field of ultra-high-energy gamma-ray astrophysics.
Nevertheless, while the LHAASO approach of absorbing the electromagnetic component of Extensive Air Showers (EAS) by burying large Water Cherenkov Detectors (WCDs) under several meters of soil~\cite{LHAASO_muon} is highly effective, it is also extremely costly and unfeasible in environmentally protected areas.

Recently a new gamma/hadron (g/h) discriminating variable, $LCm$, was proposed in~\cite{LCm}. The $LCm$ quantifies, on an event-by-event basis, the azimuthal non-uniformity in the pattern of the shower at the ground.

The asymmetries are assessed via the variable $C_{k}$, defined for each radial ring $k$ as:

\begin{equation}
C_{k} =\frac{2}{n_{k}(n_{k}-1)} 
\frac{1}{\left<S_{k}\right>}\sum_{i=1}^{n_{k}-1}\sum_{j=i+1}^{n_{k}}(S_{ik}-S_{jk})^{2}
\label{eq:CK}
\end{equation}

where $n_{k}$ is the number of stations in ring $k$, $\left<S_{k}\right>$ is the mean signal in the stations of the ring $k$, and $S_{ik}$ and $S_{jk}$ are signals in stations $i$ and $j$ of the ring $k$, respectively.

Each circular annulus $k$ is centred around the shower core position with a width of $\Delta k_r$. In this work, it is chosen as $\Delta k_r \in [10;40]\,$m, depending on the statistical power of $C_k$ profile.

The $C_k$ profile derived for each shower is then fitted through the following parameterization:

\begin{equation}
\log(C_{k}) = a  +  \frac{b}{\log \left(\frac{r_{k}}{40\,{\rm m}}\right)+1} 
\label{eq:CKfit}
\end{equation}

allowing to extract the gamma/hadron discrimination quantity, $LCm$, on an event-by-event basis, defined as $LCm \equiv \left. \log\left(  C_{k}  \right) \right|_{r_k = 360\,{\rm m}}$. 

$LCm$ has been shown to exhibit a strong correlation with the total number of muons observed at the ground, $N_{\mu}$. Furthermore, tests conducted in~\cite{LCm} on the electromagnetic ground signal suggest that $LCm$ may be capturing shower sub-structures, which are expected to be more prominent in showers dominated by hadronic interactions.  

Additionally, it has been demonstrated that $LCm$ can be generalized for use in detector arrays with varying configurations and fill factors~\cite{LCm_applications}.

Despite all these promising results, the performance of this variable was evaluated using a limited sample of EAS events, around $\mathcal{O}(10^4)$, which was adequate for energies of $100\,$TeV but insufficient to establish the necessary rejection levels at PeV energies. The next crucial step is to determine whether this discrimination power can be extended to PeV energies, which requires generating and analyzing much larger datasets of shower events (approximately one million). This is the main focus of the present article.

In this work, a strategy to simulate and handle a very large EAS sample is developed and applied to study muon-depleted proton air showers with energy deposits at the ground equivalent to PeV gamma showers. These investigations are done considering detector array configurations with different fill factors (FF), and the implications of the obtained results for the design of large ground-array gamma-ray observatories are discussed.


 \section{Simulation and large EAS sets handling}
\label{sec:simulation}

To perform the study described in the previous section, $10^6$ proton-induced showers were produced with energies between $1$ and $2\,$PeV using CORSIKA (version 7.7410)~\cite{CORSIKA}. The showers were simulated employing as hadronic interaction models for low and high energy interactions UrQMD~\cite{urqmd, urqmd2} and \qgs~\cite{qgs}, respectively. The zenith angle was fixed to $20^\circ$ with respect to the vertical, while the azimuth angle was chosen from a uniform distribution. The shower secondary particles were collected at an altitude of $4700\,$m a.s.l.\footnote{The considered altitude for this study was chosen for being the reference height for the R\&D studies being conducted by the Southern Wide-field Gamma-ray Observatory~\cite{SWGO,SWGOsim}.}

At these energies each simulated shower requires large disk space for storage,  making it impossible to store all simulations. To cope with this, two sets were extracted from the  original proton simulations : one with all the shower events below a fixed muon scale, the proton muon-depleted set -- designated throughout this paper as \emph{tail}; another, the proton reduced set -- referred to as \emph{bulk} -- with about one-hundredth of the events not selected for the first set, chosen randomly. The threshold for this decision was set to $\Nmu=5000$, where $\Nmu$ is the number of muons contained in one square kilometer. This value was verified with a smaller shower sample $\mathcal{O}(10^4)$ to be the number to select the $1\%$ of showers with the lowest number of muons.
The \emph{tail} simulation set preserves all the proton events more likely to be identified as gamma candidates if the main g/h discriminator relies on the number of muons at the ground. The latter simulation set (\emph{bulk}) is used to reconstruct the complete shape of any distribution of interest.

\begin{figure}[htb]
\centering
\includegraphics[width=0.8\linewidth]{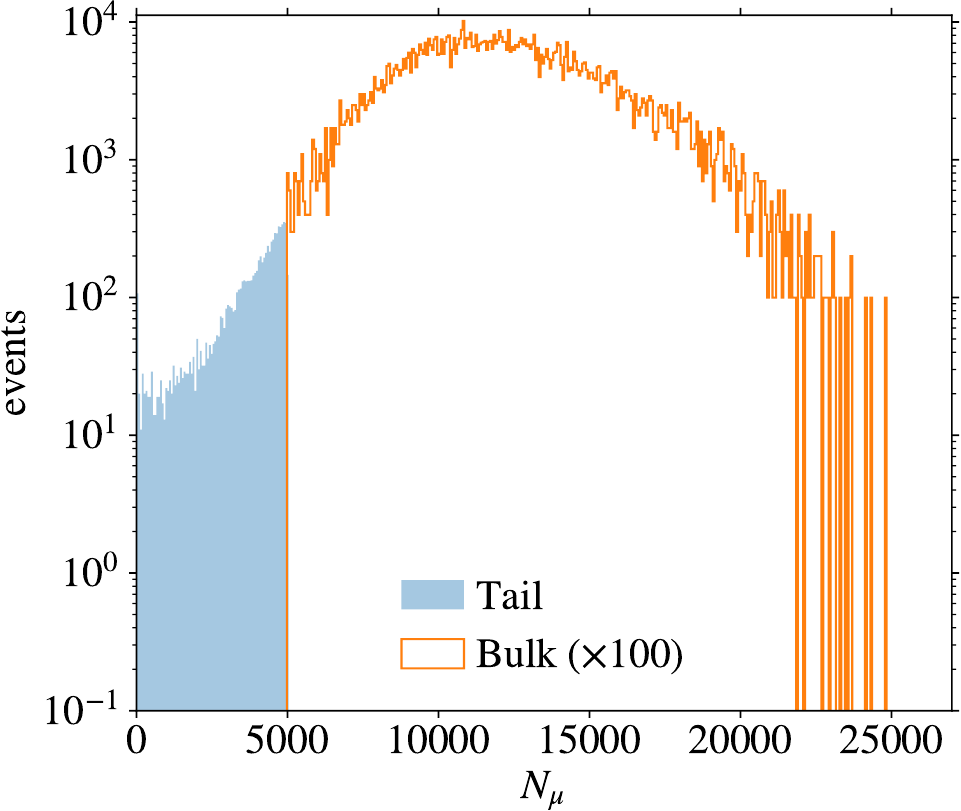}
\caption{\label{fig:Nmup} Distribution of the number of muons at the ground in the proton EAS: the blue filled bins correspond to the proton muon-depleted sample; the bins with orange contours are the proton reduced set, multiplying the mean number of the events in each bin by one hundred (the inverse of the sampling factor).}
\end{figure}

As an example, in Fig.~\ref{fig:Nmup}, the distribution of the number of muons at the ground is shown for the proton showers, putting together both sets. The size of the bin-to-bin fluctuations reflects the statistics of the corresponding samples.

\begin{figure}[htb]
\centering
\includegraphics[width=0.8\linewidth]{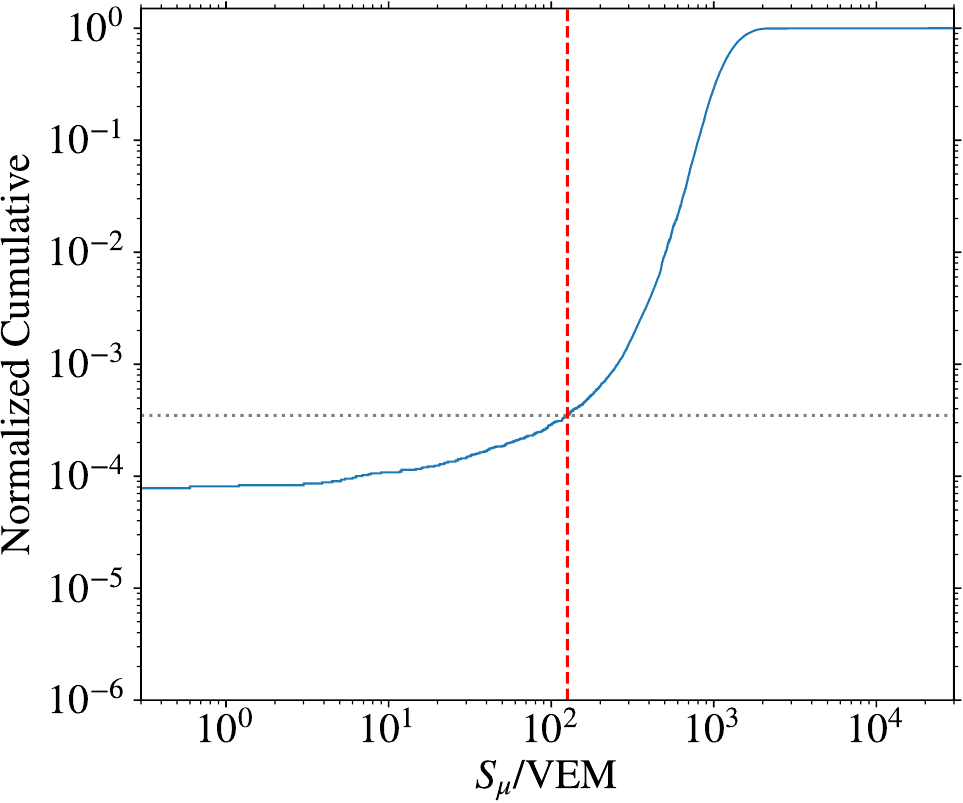}

\includegraphics[width=0.8\linewidth]{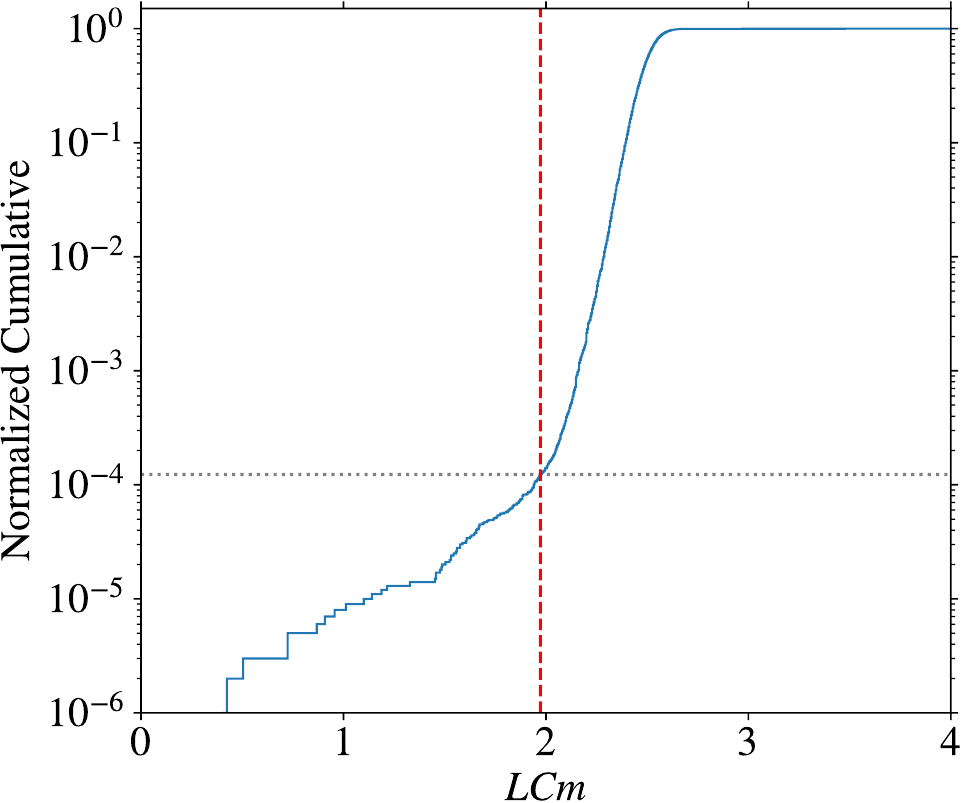}
\caption{\label{fig:Cumul}
Cumulative distributions for the $\Smu$ (top) and $LCm$ (bottom) distribution for events in the reference proton set (proton tail + proton bulk normalized to the total number of showers simulated). The red (dashed) lines define the values of $\Smu$ and $LCm$ for which the gamma set has a selection efficiency of $90\%$.}
\end{figure}

Additionally, a set of $1000$ gamma-induced showers was simulated in the same conditions described for the protons, except for the energy. The energy was fixed to $1.6\,$PeV. Such was verified to be the mean energy for which proton and gamma showers have approximately the same signal footprint at the ground for a $20^\circ$ zenith angle. It is important to note that the aim of this study is to have a reference to compare $LCm$ with $\Nmu$ and not to claim absolute background rejection power.

Following reference~\cite{LCm}, a 2D histogram with cells with an area of $\sim 12\,{\rm m^2}$ emulated a ground detector array with a fill factor equal to one (FF$=$1). Smaller FFs were obtained by masking the 2D histogram with regular patterns. A bijective correspondence between cells and the WCD stations was established, and thus, the total signal in each station is given by the sum of the expected signals due to the particles that hit the corresponding histogram cell. The amount of signal deposited by the particles in a given cell was computed through a parametrization derived using a dedicated Geant4 simulation of the water Cherenkov detector (WCD) considered in this work~\cite{Mercedes}. The parameterizations were derived for muons, electrons and protons. The latter two represent the electromagnetic and the hadronic shower component, respectively. The signal parameterizations as a function of the particle energy were built for the mean signal and its fluctuations. The fluctuations due to the stochastic processes of particle interactions and light collection and the fluctuations of the muon tracklengths in the station were considered.

\section{Gamma/hadron discrimination}
\label{sec:g-h}

In this work, the experimental proxy to $\Nmu$ is the total amount of signal recorded by the WCDs due to the passage of muons, $\Smu$. The quantity $\Smu$ is expressed in Vertical Equivalent Muon (VEM) units, representing the number of photoelectrons recorded by the WCD photosensor, normalized to the signal produced by a vertically-centered muon passing through the center of the WCD~\cite{VEM,PierreAugerObservatory}.
It is assumed that $\Smu$ can be obtained without any uncertainty other than the signal and tracklengths fluctuations mentioned before.

In Fig.~\ref{fig:Cumul}, the cumulative distributions of the $\Smu$ (top) and $LCm$ (bottom) variables, obtained assuming a detector array with a fill factor of $12.5\%$, are shown.

To evaluate the g/h discrimination power of the probed quantities, we examined the number of events that survive after applying a cut on these quantities, ensuring that $90\%$ of the gamma-simulated events survive. Throughout this work, the fraction of events below these cuts (proton selection efficiencies) will be referred to as $S^g_{\mu}$ for the recorded muon signal and $LCm^g$ for the reconstructed $LCm$ value, respectively. Note that the ultimate goal of a gamma-ray observatory is to achieve high purity in gamma-induced shower detection, which translates to a high proton rejection efficiency.

From Fig.~\ref{fig:Cumul}, it is observed that the values corresponding to a 90\% gamma shower selection efficiency in each of these cumulative distributions are $S^g_{\mu}= 4.29 \times 10^{-4}$ and $LCm^g = 1.39 \times 10^{-4}$, respectively. Consequently, the $LCm$ has a lower residual background of protons for selecting gamma showers, approximately a factor of $3$ with respect to $\Smu$.

The same study was conducted assuming a sparser array with FF$=1.4\%$. The proton selection efficiencies become now: $S^{g}_{\mu}= 9.33 \times 10^{-4}$ and $LCm^{g}= 6.10 \times 10^{-4}$, making $LCm$ a slightly better discriminator ($\sim 50\%)$.

Again, we note that the above numbers should be compared only in relative terms. The evaluation of the absolute value of $S^g_{\mu}$ and $LCm^g$ would require fully reconstructed shower events, which necessitates a much larger dataset.

\section{$LCm$-$N_{\mu}$ correlations}
\label{sec:analysis}

In this section, the correlation between the observed number of muons at the ground and the  $LCm$ variable is discussed, focusing on the ability to distinguish PeV gamma-induced showers from the cosmic-ray background.

Shown in Fig.~\ref{fig:NmuLCm012}, is the observed $LCm$-$S_{\mu}$ correlation for the considered samples, assuming a detector array with FF$=12.5\%$. Shower events with $\Smu = 0$ were placed at the extreme left of the plot, while events with poor quality\footnote{The criterion was to require that the $C_k$ profile, constructed with radial bins of $30\,$m,  had more than two degrees of freedom in the fit used to extract $LCm$~\cite{LCm}.} are displayed at the top.


The lines indicate the values of $S^g_{\mu}$ and $LCm^g$, defined in the previous section (see Fig.~\ref{fig:Cumul}), which delimit the regions that preserve 90$\%$ of the gamma events. These lines define four areas of interest:

\begin{figure}[!t]
\centering
\includegraphics[width=0.99\linewidth]{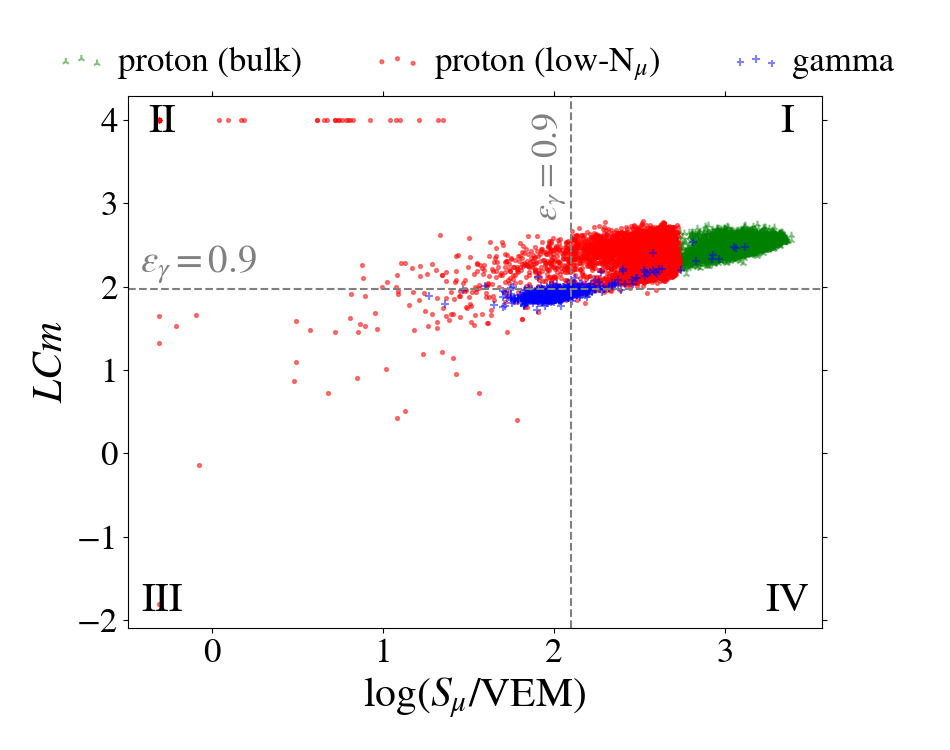}
\caption{\label{fig:NmuLCm012} Correlation between $\log(\Smu)$ and $LCm$ for the muon-depleted (red), proton bulk (green) and gamma (blue) events. The dashed grey lines indicate the cuts on $\Nmu$ and $LCm$ to select $90\%$ for the gamma showers. The discrimination quantities were computed assuming a detector array with a fill factor of $12.5\%$.}
\end{figure}

\begin{itemize}

\item Region I - $S_{\mu} > S^g_{\mu}$ and $LCm > LCm^g$: events rejected when using either $LCm$ or $S_{\mu}$ as the g/h discriminator;

\item Region II - $S_{\mu} < S^g_{\mu}$ and $LCm > LCm^g$: events accepted when using $S_{\mu}$ as the g/h discriminator but rejected when using $LCm$ as the discriminator;

\item Region III - $S_{\mu} < S^g_{\mu}$ and $LCm < LCm^g$: events accepted using either $LCm$ or $S_{\mu}$ as the g/h discriminator;

\item Region IV - $S_{\mu} > S^g_{\mu}$ and $LCm < LCm^g$: events accepted when using $LCm$ as the g/h discriminator but rejected when using $S_{\mu}$ as the discriminator.

 \end{itemize}

Considering the total simulated statistics of $10^6$ proton showers, the fraction of events that would be in each of these regions, assuming FF$=12.5\%$ are: Region I - $9.99 \times 10^{-1}$; Region II - $4.03 \times 10^{-4}$; Region III - $9.90 \times 10^{-5}$; Region IV - $1.00 \times 10^{-5}$.

\begin{figure}[!t]
\centering
\includegraphics[width=0.99\linewidth]{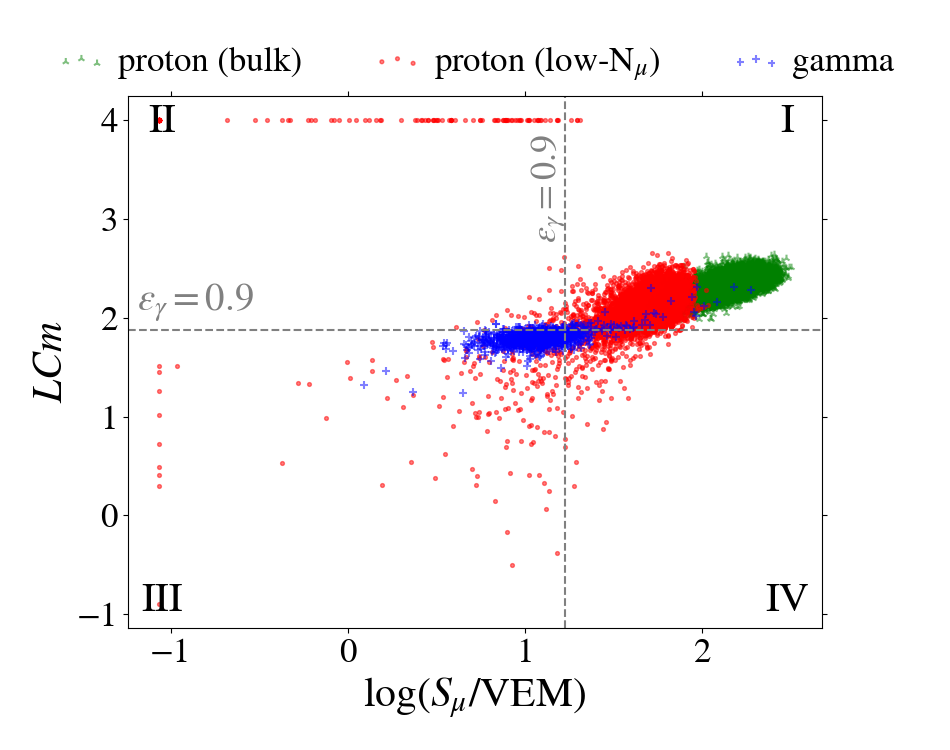}
\caption{\label{fig:NmuLCmFF0014} Same plot as the one displayed in Fig.~\ref{fig:NmuLCm012} but assuming a detector array with a fill factor of $1.4\%$.}
\end{figure}

A low FF is mandatory for a real detector array with a size able to collect useful event statistics at the PeV energies. In these terms, the previous figures were redone considering now FF$=1.4\%$ (Fig.~\ref{fig:NmuLCmFF0014}). The fraction of events that would be in each of the above-defined regions are now:
Region I - $9.99 \times 10^{-1}$; Region II - $4.09 \times 10^{-4}$; Region III - $3.52 \times 10^{-4}$; Region IV - $1.15 \times 10^{-4}$.

The potential impact of a signal threshold due to the station triggering probability was also investigated. The threshold was set as high as 10 photoelectrons~\cite{Mercedes} with no visible effect on the analysis.

Although it is beyond the scope of this paper, we would like to emphasize the high correlation between $\log(\Smu)$ and $LCm$, which could potentially be explored to probe the shower muon content without the need for dedicated muon counters.

Additionally, for all tested fill factors, the number of events in Region II is higher than the number of events in Region IV, implying that the shower can be discriminated through the azimuthal fluctuations even if the number of muons is compatible with those corresponding to a gamma primary with equivalent energy. This likely indicates that the electromagnetic shower component still retains information about the nature of the primary particle. In fact, in~\cite{LCm}, it was shown that at 100~TeV $LCm$ attains discrimination power even if only the electromagnetic shower component is considered. The result obtained in the present work extends the confirmation of this interesting feature for the rare muon-depleted showers that constitute the primary background for accurately identifying showers at PeV energies.

Finally, one should note that this study used the quantity $\Smu$ as a proxy for $\Nmu$ and it might be argued that a detector other than a WCD might lead to different conclusions. To test this, $LCm$ was computed for an array with FF$=12.5\%$ and directly compared to the total number of muons at the ground in $1\,{\rm km^2}$ (FF$=100\%$). In these conditions, unfeasible for a realistic experiment, the discrimination capability of $LCm$ was verified to continue to surpass those of $\Nmu$ by a factor of $5$.

\section{Discussion and conclusions}
\label{sec:conclusions}

The number of muons produced on average in a high energy hadronic-induced shower that reaches the ground at a high altitude is an order of magnitude higher than that produced in a gamma-induced shower of the same reconstructed energy. Thus, $N_{\mu}$ is an excellent g/h discriminator,  ensuring rejection levels of the order of $10^{-4}$ at the PeV energies~\cite{LHAASO_KM2A_performance}. However, at these energies and altitudes, the number of EAS photons and electrons reaching the ground is many orders of magnitude higher than the number of their companion muons.
In this way, directly counting muons requires the use of shielded detectors with some inert material such as earth (e.g. \cite{1994NIMPA,Aab_2021,LHAASO_muon}), water  (e.g. \cite{MARTA}, \cite{DLWCD_Antoine}) or concrete and iron (e.g. \cite{AGASA}, \cite{KASCADE}). It is an effective strategy, but highly costly to implement in large-area observatories (approximately a few ${\rm km^2}$). 

In this work, a simulation strategy was conceived to analyse the rare muon-depleted shower events, the main background source for gamma PeV showers. With it, it was shown that the $\Smu$ and $LCm$ variables continue to have a high correlation, outperforming the discrimination power of the  direct $N_{\mu}$-based methods. This conclusion holds for all the tested array fill factors, which span from $100\%$ down to $1.4\%$ and it is valid even when an ideal muon detector with perfect efficiency is assumed, while the LCm calculations rely on realistic simulations of the WCD signals recorded by  stations on the ground surface.

The findings in this work further support the use of $LCm$ as an excellent gamma-hadron discriminator, emphasizing its potential as a valuable tool for future ground-based, wide field-of-view gamma-ray experiments targeting the hundreds of TeV to PeV energy range~\cite{SWGO}.

\section*{Acknowledgments}
We would like to thank Ulisses Barres de Almeida for carefully reading the manuscript and for useful comments.
The authors also thank for the financial support by OE - Portugal, FCT, I. P., under project PTDC/FIS-PAR/4300/2020 and \url{https://doi.org/10.54499/2024.06879.CERN}, 
MEYS of the Czech Republic under
project FORTE No. CZ.02.01.01/00/22\_008/0004632, co-funded by the European Union,
and Czech Science Foundation under grant 23-05827S. R.~C.\ is grateful for the financial support by OE - Portugal, FCT, I. P., under DL57/2016/cP1330/cT0002. L.~G. is grateful for the financial support by FCT under PRT/BD/154192/2022.

\bibliography{references}

\begin{thebibliography}{22}%
\makeatletter
\providecommand \@ifxundefined [1]{%
 \@ifx{#1\undefined}
}%
\providecommand \@ifnum [1]{%
 \ifnum #1\expandafter \@firstoftwo
 \else \expandafter \@secondoftwo
 \fi
}%
\providecommand \@ifx [1]{%
 \ifx #1\expandafter \@firstoftwo
 \else \expandafter \@secondoftwo
 \fi
}%
\providecommand \natexlab [1]{#1}%
\providecommand \enquote  [1]{``#1''}%
\providecommand \bibnamefont  [1]{#1}%
\providecommand \bibfnamefont [1]{#1}%
\providecommand \citenamefont [1]{#1}%
\providecommand \href@noop [0]{\@secondoftwo}%
\providecommand \href [0]{\begingroup \@sanitize@url \@href}%
\providecommand \@href[1]{\@@startlink{#1}\@@href}%
\providecommand \@@href[1]{\endgroup#1\@@endlink}%
\providecommand \@sanitize@url [0]{\catcode `\\12\catcode `\$12\catcode
  `\&12\catcode `\#12\catcode `\^12\catcode `\_12\catcode `\%12\relax}%
\providecommand \@@startlink[1]{}%
\providecommand \@@endlink[0]{}%
\providecommand \url  [0]{\begingroup\@sanitize@url \@url }%
\providecommand \@url [1]{\endgroup\@href {#1}{\urlprefix }}%
\providecommand \urlprefix  [0]{URL }%
\providecommand \Eprint [0]{\href }%
\providecommand \doibase [0]{https://doi.org/}%
\providecommand \selectlanguage [0]{\@gobble}%
\providecommand \bibinfo  [0]{\@secondoftwo}%
\providecommand \bibfield  [0]{\@secondoftwo}%
\providecommand \translation [1]{[#1]}%
\providecommand \BibitemOpen [0]{}%
\providecommand \bibitemStop [0]{}%
\providecommand \bibitemNoStop [0]{.\EOS\space}%
\providecommand \EOS [0]{\spacefactor3000\relax}%
\providecommand \BibitemShut  [1]{\csname bibitem#1\endcsname}%
\let\auto@bib@innerbib\@empty
\bibitem [{\citenamefont {Cao}\ \emph {et~al.}(2021)\citenamefont {Cao} \emph
  {et~al.}}]{LHAASO_PeV}%
  \BibitemOpen
  \bibfield  {author} {\bibinfo {author} {\bibfnamefont {Z.}~\bibnamefont
  {Cao}} \emph {et~al.},\ }\bibfield  {title} {\bibinfo {title}
  {{Ultrahigh-energy photons up to 1.4 petaelectronvolts from 12 $\gamma$ ray
  Galactic sources}},\ }\href {https://doi.org/10.1038/s41586-021-03498-z}
  {\bibfield  {journal} {\bibinfo  {journal} {\it Nature}\ }\textbf {\bibinfo
  {volume} {594}},\ \bibinfo {pages} {33} (\bibinfo {year} {2021})}\BibitemShut
  {NoStop}%
\bibitem [{\citenamefont {Zuo}\ \emph {et~al.}(2015)\citenamefont {Zuo} \emph
  {et~al.}}]{LHAASO_muon}%
  \BibitemOpen
  \bibfield  {author} {\bibinfo {author} {\bibfnamefont {X.}~\bibnamefont
  {Zuo}} \emph {et~al.} (\bibinfo {collaboration} {LHAASO}),\ }\bibfield
  {title} {\bibinfo {title} {{Design and performances of prototype muon
  detectors of LHAASO-KM2A}},\ }\href
  {https://doi.org/10.1016/j.nima.2015.04.010} {\bibfield  {journal} {\bibinfo
  {journal} {Nucl. Instrum. Meth. A}\ }\textbf {\bibinfo {volume} {789}},\
  \bibinfo {pages} {143} (\bibinfo {year} {2015})}\BibitemShut {NoStop}%
\bibitem [{\citenamefont {Concei\c{c}\~ao}\ \emph {et~al.}(2022)\citenamefont
  {Concei\c{c}\~ao}, \citenamefont {Gibilisco}, \citenamefont {Pimenta},\ and\
  \citenamefont {Tom\'e}}]{LCm}%
  \BibitemOpen
  \bibfield  {author} {\bibinfo {author} {\bibfnamefont {R.}~\bibnamefont
  {Concei\c{c}\~ao}}, \bibinfo {author} {\bibfnamefont {L.}~\bibnamefont
  {Gibilisco}}, \bibinfo {author} {\bibfnamefont {M.}~\bibnamefont {Pimenta}},\
  and\ \bibinfo {author} {\bibfnamefont {B.}~\bibnamefont {Tom\'e}},\
  }\bibfield  {title} {\bibinfo {title} {{Gamma/hadron discrimination at high
  energies through the azimuthal fluctuations of air shower particle
  distributions at the ground}},\ }\href
  {https://doi.org/10.1088/1475-7516/2022/10/086} {\bibfield  {journal}
  {\bibinfo  {journal} {JCAP}\ }\textbf {\bibinfo {volume} {10}},\ \bibinfo
  {pages} {086}},\ \Eprint {https://arxiv.org/abs/2204.12337} {arXiv:2204.12337
  [hep-ph]} \BibitemShut {NoStop}%
\bibitem [{\citenamefont {{Concei\c{c}\~{a}o}}\ \emph
  {et~al.}(2023)\citenamefont {{Concei\c{c}\~{a}o}}, \citenamefont {Costa},
  \citenamefont {Gibilisco}, \citenamefont {Pimenta},\ and\ \citenamefont
  {Tome}}]{LCm_applications}%
  \BibitemOpen
  \bibfield  {author} {\bibinfo {author} {\bibfnamefont {R.}~\bibnamefont
  {{Concei\c{c}\~{a}o}}}, \bibinfo {author} {\bibfnamefont {P.~J.}\
  \bibnamefont {Costa}}, \bibinfo {author} {\bibfnamefont {L.}~\bibnamefont
  {Gibilisco}}, \bibinfo {author} {\bibfnamefont {M.}~\bibnamefont {Pimenta}},\
  and\ \bibinfo {author} {\bibfnamefont {B.}~\bibnamefont {Tome}},\ }\bibfield
  {title} {\bibinfo {title} {{The gamma/hadron discriminator LCm in realistic
  air shower array experiments}},\ }\href
  {https://doi.org/10.1140/epjc/s10052-023-12106-5} {\bibfield  {journal}
  {\bibinfo  {journal} {Eur. Phys. J. C}\ }\textbf {\bibinfo {volume} {83}},\
  \bibinfo {pages} {932} (\bibinfo {year} {2023})},\ \Eprint
  {https://arxiv.org/abs/2304.05348} {arXiv:2304.05348 [astro-ph.HE]}
  \BibitemShut {NoStop}%
\bibitem [{\citenamefont {Heck}\ \emph {et~al.}(1998)\citenamefont {Heck},
  \citenamefont {Knapp}, \citenamefont {Capdevielle}, \citenamefont {Schatz},\
  and\ \citenamefont {Thouw}}]{CORSIKA}%
  \BibitemOpen
  \bibfield  {author} {\bibinfo {author} {\bibfnamefont {D.}~\bibnamefont
  {Heck}}, \bibinfo {author} {\bibfnamefont {J.}~\bibnamefont {Knapp}},
  \bibinfo {author} {\bibfnamefont {J.}~\bibnamefont {Capdevielle}}, \bibinfo
  {author} {\bibfnamefont {G.}~\bibnamefont {Schatz}},\ and\ \bibinfo {author}
  {\bibfnamefont {T.}~\bibnamefont {Thouw}},\ }\bibfield  {title} {\bibinfo
  {title} {A monte carlo code to simulate extensive air showers},\ }\href@noop
  {} {\bibfield  {journal} {\bibinfo  {journal} {Report FZKA}\ }\textbf
  {\bibinfo {volume} {6019}} (\bibinfo {year} {1998})}\BibitemShut {NoStop}%
\bibitem [{\citenamefont {Bass}\ \emph {et~al.}(1998)\citenamefont {Bass} \emph
  {et~al.}}]{urqmd}%
  \BibitemOpen
  \bibfield  {author} {\bibinfo {author} {\bibfnamefont {S.~A.}\ \bibnamefont
  {Bass}} \emph {et~al.},\ }\bibfield  {title} {\bibinfo {title} {{Microscopic
  models for ultrarelativistic heavy ion collisions}},\ }\href
  {https://doi.org/10.1016/S0146-6410(98)00058-1} {\bibfield  {journal}
  {\bibinfo  {journal} {Prog. Part. Nucl. Phys.}\ }\textbf {\bibinfo {volume}
  {41}},\ \bibinfo {pages} {255} (\bibinfo {year} {1998})},\ \Eprint
  {https://arxiv.org/abs/nucl-th/9803035} {arXiv:nucl-th/9803035} \BibitemShut
  {NoStop}%
\bibitem [{\citenamefont {Bleicher}\ \emph {et~al.}(1999)\citenamefont
  {Bleicher} \emph {et~al.}}]{urqmd2}%
  \BibitemOpen
  \bibfield  {author} {\bibinfo {author} {\bibfnamefont {M.}~\bibnamefont
  {Bleicher}} \emph {et~al.},\ }\bibfield  {title} {\bibinfo {title}
  {{Relativistic hadron hadron collisions in the ultrarelativistic quantum
  molecular dynamics model}},\ }\href
  {https://doi.org/10.1088/0954-3899/25/9/308} {\bibfield  {journal} {\bibinfo
  {journal} {J. Phys. G}\ }\textbf {\bibinfo {volume} {25}},\ \bibinfo {pages}
  {1859} (\bibinfo {year} {1999})},\ \Eprint
  {https://arxiv.org/abs/hep-ph/9909407} {arXiv:hep-ph/9909407} \BibitemShut
  {NoStop}%
\bibitem [{\citenamefont {Ostapchenko}(2011)}]{qgs}%
  \BibitemOpen
  \bibfield  {author} {\bibinfo {author} {\bibfnamefont {S.}~\bibnamefont
  {Ostapchenko}},\ }\bibfield  {title} {\bibinfo {title} {{Monte Carlo
  treatment of hadronic interactions in enhanced Pomeron scheme: I. QGSJET-II
  model}},\ }\href {https://doi.org/10.1103/PhysRevD.83.014018} {\bibfield
  {journal} {\bibinfo  {journal} {Phys. Rev. D}\ }\textbf {\bibinfo {volume}
  {83}},\ \bibinfo {pages} {014018} (\bibinfo {year} {2011})},\ \Eprint
  {https://arxiv.org/abs/1010.1869} {arXiv:1010.1869 [hep-ph]} \BibitemShut
  {NoStop}%
\bibitem [{Note1()}]{Note1}%
  \BibitemOpen
  \bibinfo {note} {The considered altitude for this study was chosen for being
  the reference height for the R\&D studies being conducted by the Southern
  Wide-field Gamma-ray Observatory~\cite {SWGO,SWGOsim}.}\BibitemShut {Stop}%
\bibitem [{\citenamefont {Assis}\ \emph {et~al.}(2022)\citenamefont {Assis}
  \emph {et~al.}}]{Mercedes}%
  \BibitemOpen
  \bibfield  {author} {\bibinfo {author} {\bibfnamefont {P.}~\bibnamefont
  {Assis}} \emph {et~al.},\ }\bibfield  {title} {\bibinfo {title} {{The
  Mercedes water Cherenkov detector}},\ }\href
  {https://doi.org/10.1140/epjc/s10052-022-10857-1} {\bibfield  {journal}
  {\bibinfo  {journal} {Eur. Phys. J. C}\ }\textbf {\bibinfo {volume} {82}},\
  \bibinfo {pages} {899} (\bibinfo {year} {2022})},\ \Eprint
  {https://arxiv.org/abs/2203.08782} {arXiv:2203.08782 [physics.ins-det]}
  \BibitemShut {NoStop}%
\bibitem [{\citenamefont {Bertou}\ \emph {et~al.}(2006)\citenamefont {Bertou}
  \emph {et~al.}}]{VEM}%
  \BibitemOpen
  \bibfield  {author} {\bibinfo {author} {\bibfnamefont {X.}~\bibnamefont
  {Bertou}} \emph {et~al.} (\bibinfo {collaboration} {Pierre Auger}),\
  }\bibfield  {title} {\bibinfo {title} {{Calibration of the surface array of
  the Pierre Auger Observatory}},\ }\href
  {https://doi.org/10.1016/j.nima.2006.07.066} {\bibfield  {journal} {\bibinfo
  {journal} {Nucl. Instrum. Meth. A}\ }\textbf {\bibinfo {volume} {568}},\
  \bibinfo {pages} {839} (\bibinfo {year} {2006})},\ \Eprint
  {https://arxiv.org/abs/2102.01656} {arXiv:2102.01656 [astro-ph.HE]}
  \BibitemShut {NoStop}%
\bibitem [{\citenamefont {Aab}\ \emph {et~al.}(2015)\citenamefont {Aab} \emph
  {et~al.}}]{PierreAugerObservatory}%
  \BibitemOpen
  \bibfield  {author} {\bibinfo {author} {\bibfnamefont {A.}~\bibnamefont
  {Aab}} \emph {et~al.} (\bibinfo {collaboration} {Pierre Auger}),\ }\bibfield
  {title} {\bibinfo {title} {{The Pierre Auger Cosmic Ray Observatory}},\
  }\href {https://doi.org/10.1016/j.nima.2015.06.058} {\bibfield  {journal}
  {\bibinfo  {journal} {Nucl. Instrum. Meth. A}\ }\textbf {\bibinfo {volume}
  {798}},\ \bibinfo {pages} {172} (\bibinfo {year} {2015})},\ \Eprint
  {https://arxiv.org/abs/1502.01323} {arXiv:1502.01323 [astro-ph.IM]}
  \BibitemShut {NoStop}%
\bibitem [{Note2()}]{Note2}%
  \BibitemOpen
  \bibinfo {note} {The criterion was to require that the $C_k$ profile,
  constructed with radial bins of $30\protect \,$m, had more than two degrees
  of freedom in the fit used to extract $LCm$~\cite {LCm}.}\BibitemShut {Stop}%
\bibitem [{\citenamefont {Aharonian}\ \emph {et~al.}(2021)\citenamefont
  {Aharonian} \emph {et~al.}}]{LHAASO_KM2A_performance}%
  \BibitemOpen
  \bibfield  {author} {\bibinfo {author} {\bibfnamefont {F.}~\bibnamefont
  {Aharonian}} \emph {et~al.},\ }\bibfield  {title} {\bibinfo {title} {{The
  observation of the Crab Nebula with LHAASO-KM2A for the performance study}},\
  }\href {https://doi.org/10.1088/1674-1137/abd01b} {\bibfield  {journal}
  {\bibinfo  {journal} {Chin. Phys. C}\ }\textbf {\bibinfo {volume} {45}},\
  \bibinfo {pages} {025002} (\bibinfo {year} {2021})},\ \Eprint
  {https://arxiv.org/abs/2010.06205} {arXiv:2010.06205 [astro-ph.HE]}
  \BibitemShut {NoStop}%
\bibitem [{\citenamefont {{Borione}}\ and\ \citenamefont {et.
  al}(1994)}]{1994NIMPA}%
  \BibitemOpen
  \bibfield  {author} {\bibinfo {author} {\bibfnamefont {A.}~\bibnamefont
  {{Borione}}}\ and\ \bibinfo {author} {\bibnamefont {et. al}},\ }\bibfield
  {title} {\bibinfo {title} {{A large air shower array to search for
  astrophysical sources emitting {\ensuremath{\gamma}}-rays with energies $\ge$
  10$^{14}$ eV}},\ }\href {https://doi.org/10.1016/0168-9002(94)90722-6}
  {\bibfield  {journal} {\bibinfo  {journal} {Nuclear Instruments and Methods
  in Physics Research A}\ }\textbf {\bibinfo {volume} {346}},\ \bibinfo {pages}
  {329} (\bibinfo {year} {1994})}\BibitemShut {NoStop}%
\bibitem [{\citenamefont {Aab}\ and\ \citenamefont {et. al}(2021)}]{Aab_2021}%
  \BibitemOpen
  \bibfield  {author} {\bibinfo {author} {\bibfnamefont {A.}~\bibnamefont
  {Aab}}\ and\ \bibinfo {author} {\bibnamefont {et. al}},\ }\bibfield  {title}
  {\bibinfo {title} {Calibration of the underground muon detector of the
  {P}ierre {A}uger {O}bservatory},\ }\href
  {https://doi.org/10.1088/1748-0221/16/04/p04003} {\bibfield  {journal}
  {\bibinfo  {journal} {Journal of Instrumentation}\ }\textbf {\bibinfo
  {volume} {16}}\bibinfo  {number} { (04)},\ \bibinfo {pages}
  {P04003}}\BibitemShut {NoStop}%
\bibitem [{\citenamefont {Abreu}\ \emph {et~al.}(2018)\citenamefont {Abreu}
  \emph {et~al.}}]{MARTA}%
  \BibitemOpen
\bibfield  {number} {  }\bibfield  {author} {\bibinfo {author} {\bibfnamefont
  {P.}~\bibnamefont {Abreu}} \emph {et~al.},\ }\bibfield  {title} {\bibinfo
  {title} {{MARTA: a high-energy cosmic-ray detector concept for high-accuracy
  muon measurement}},\ }\href {https://doi.org/10.1140/epjc/s10052-018-5820-2}
  {\bibfield  {journal} {\bibinfo  {journal} {Eur. Phys. J. C}\ }\textbf
  {\bibinfo {volume} {78}},\ \bibinfo {pages} {333} (\bibinfo {year} {2018})},\
  \Eprint {https://arxiv.org/abs/1712.07685} {arXiv:1712.07685
  [physics.ins-det]} \BibitemShut {NoStop}%
\bibitem [{\citenamefont {Letessier-Selvon}\ \emph {et~al.}(2014)\citenamefont
  {Letessier-Selvon}, \citenamefont {Billoir}, \citenamefont {Blanco},
  \citenamefont {Mari\c{s}},\ and\ \citenamefont {Settimo}}]{DLWCD_Antoine}%
  \BibitemOpen
  \bibfield  {author} {\bibinfo {author} {\bibfnamefont {A.}~\bibnamefont
  {Letessier-Selvon}}, \bibinfo {author} {\bibfnamefont {P.}~\bibnamefont
  {Billoir}}, \bibinfo {author} {\bibfnamefont {M.}~\bibnamefont {Blanco}},
  \bibinfo {author} {\bibfnamefont {I.~C.}\ \bibnamefont {Mari\c{s}}},\ and\
  \bibinfo {author} {\bibfnamefont {M.}~\bibnamefont {Settimo}},\ }\bibfield
  {title} {\bibinfo {title} {{Layered water Cherenkov detector for the study of
  ultra high energy cosmic rays}},\ }\href
  {https://doi.org/10.1016/j.nima.2014.08.029} {\bibfield  {journal} {\bibinfo
  {journal} {Nucl. Instrum. Meth. A}\ }\textbf {\bibinfo {volume} {767}},\
  \bibinfo {pages} {41} (\bibinfo {year} {2014})},\ \Eprint
  {https://arxiv.org/abs/1405.5699} {arXiv:1405.5699 [astro-ph.IM]}
  \BibitemShut {NoStop}%
\bibitem [{\citenamefont {Chiba}\ \emph {et~al.}(1992)\citenamefont {Chiba}
  \emph {et~al.}}]{AGASA}%
  \BibitemOpen
  \bibfield  {author} {\bibinfo {author} {\bibfnamefont {N.}~\bibnamefont
  {Chiba}} \emph {et~al.},\ }\bibfield  {title} {\bibinfo {title} {{Akeno giant
  air shower array (AGASA) covering 100-km**2 area}},\ }\href
  {https://doi.org/10.1016/0168-9002(92)90882-5} {\bibfield  {journal}
  {\bibinfo  {journal} {Nucl. Instrum. Meth. A}\ }\textbf {\bibinfo {volume}
  {311}},\ \bibinfo {pages} {338} (\bibinfo {year} {1992})}\BibitemShut
  {NoStop}%
\bibitem [{\citenamefont {Bozdog}\ \emph {et~al.}(2001)\citenamefont {Bozdog}
  \emph {et~al.}}]{KASCADE}%
  \BibitemOpen
  \bibfield  {author} {\bibinfo {author} {\bibfnamefont {H.}~\bibnamefont
  {Bozdog}} \emph {et~al.},\ }\bibfield  {title} {\bibinfo {title} {{The
  detector system for measurement of multiple cosmic muons in the central
  detector of KASCADE}},\ }\href
  {https://doi.org/10.1016/S0168-9002(01)00673-8} {\bibfield  {journal}
  {\bibinfo  {journal} {Nucl. Instrum. Meth. A}\ }\textbf {\bibinfo {volume}
  {465}},\ \bibinfo {pages} {455} (\bibinfo {year} {2001})}\BibitemShut
  {NoStop}%
\bibitem [{\citenamefont {Concei\c{c}\~ao}(2023)}]{SWGO}%
  \BibitemOpen
  \bibfield  {author} {\bibinfo {author} {\bibfnamefont {R.}~\bibnamefont
  {Concei\c{c}\~ao}} (\bibinfo {collaboration} {SWGO}),\ }\bibfield  {title}
  {\bibinfo {title} {{The Southern Wide-field Gamma-ray Observatory}},\ }\href
  {https://doi.org/10.22323/1.444.0963} {\bibfield  {journal} {\bibinfo
  {journal} {PoS}\ }\textbf {\bibinfo {volume} {ICRC2023}},\ \bibinfo {pages}
  {963} (\bibinfo {year} {2023})},\ \Eprint {https://arxiv.org/abs/2309.04577}
  {arXiv:2309.04577 [hep-ex]} \BibitemShut {NoStop}%
\bibitem [{\citenamefont {Schoorlemmer}\ \emph {et~al.}(2021)\citenamefont
  {Schoorlemmer}, \citenamefont {Concei\c{c}\~ao},\ and\ \citenamefont
  {Smith}}]{SWGOsim}%
  \BibitemOpen
  \bibfield  {author} {\bibinfo {author} {\bibfnamefont {H.}~\bibnamefont
  {Schoorlemmer}}, \bibinfo {author} {\bibfnamefont {R.}~\bibnamefont
  {Concei\c{c}\~ao}},\ and\ \bibinfo {author} {\bibfnamefont {A.~J.}\
  \bibnamefont {Smith}},\ }\bibfield  {title} {\bibinfo {title} {{Simulating
  the performance of the Southern Wide-view Gamma-ray Observatory}},\ }\href
  {https://doi.org/10.22323/1.395.0903} {\bibfield  {journal} {\bibinfo
  {journal} {PoS}\ }\textbf {\bibinfo {volume} {ICRC2021}},\ \bibinfo {pages}
  {903} (\bibinfo {year} {2021})}\BibitemShut {NoStop}%
\end{thebibliography}%


\end{document}